\documentclass[aps,prl,twocolumn,superscriptaddress,groupedaddress,scrartcl]{revtex4}  
\usepackage{graphicx}  
\usepackage{dcolumn}   
\usepackage{bm}        
\usepackage{amssymb}   
\usepackage{float}
\usepackage{siunitx}
\usepackage[version=3]{mhchem}
\usepackage{comment}
\begin{document}


\title{The observation of photon echoes from evanescently coupled rare-earth ions in a planar waveguide}

\author{Sara Marzban} 
\address{Centre for Quantum Computation and Communication Technology, Laser Physics Centre, The Australian National University, Canberra, Australian Capital Territory 0200, Australia}
\author{John G. Bartholomew}
\address{Centre for Quantum Computation and Communication Technology, Laser Physics Centre, The Australian National University, Canberra, Australian Capital Territory 0200, Australia}
\author{Stephen Madden}
\address{Laser Physics Centre, The Australian National University, Canberra, Australian Capital Territory 0200, Australia}
\author{ Khu Vu}
\address{Laser Physics Centre, The Australian National University, Canberra, Australian Capital Territory 0200, Australia}
\author{Matthew J. Sellars}
\address{Centre for Quantum Computation and Communication Technology, Laser Physics Centre, The Australian National University, Canberra, Australian Capital Territory 0200, Australia}       
\date{\today}

\begin{abstract}
We report the measurement of the inhomogeneous linewidth, homogeneous linewidth and spin state lifetime of Pr$^{3+}$ ions in a novel waveguide architecture. The TeO$_{2}$ slab waveguide deposited on a bulk Pr$^{3+}$:Y$_{2}$SiO$_{5}$ crystal allows the  $^3$H$_4 \leftrightarrow ^1$D$_2$ transition of Pr$^{3+}$ ions to be probed by the optical evanescent field that extends into the substrate. The 2 GHz inhomogeneous linewidth, the optical coherence time of $70\pm5$ $\mu$s, and the spin state lifetime of $9.8\pm0.3$ s indicate that the properties of ions interacting with the waveguide mode are consistent with those of bulk ions. This result establishes the foundation for large, integrated and high performance rare-earth-ion quantum systems based on a waveguide platform.
\end{abstract}

\maketitle
Ensembles of rare-earth-ion optical centers in transparent crystals are an appealing system for realizing integrated quantum hardware. Within this single solid-state system there is the potential to combine the generation and manipulation of flying qubits, such as photons, for state distribution with stationary qubits for long term quantum state storage. Several important components for such an integrated quantum system have been demonstrated in rare-earth-ion materials, such as efficient quantum memories for light~\cite{Hedges2010, Jobez2014} and non-classical photon sources~\cite{Ledingham2012}. Furthermore, dynamic filtering~\cite{Beavan2013}, signal modulation~\cite{Tay2010}, pulse sequencing~\cite{Saglamyurek2014} and beam routing~\cite{Wang2009b} can also be performed by harnessing the ion-light interactions in rare-earth systems. To progress from these lab-based proof-of-principle demonstrations of individual components to a system that integrates a large number of active and passive devices on a single crystal chip, a scalable architecture that preserves the current material properties is required.

\begin{figure}
\begin{center}
\includegraphics[width=0.45\textwidth, trim = 1.6cm 7cm 7cm 0.8cm, clip = true]{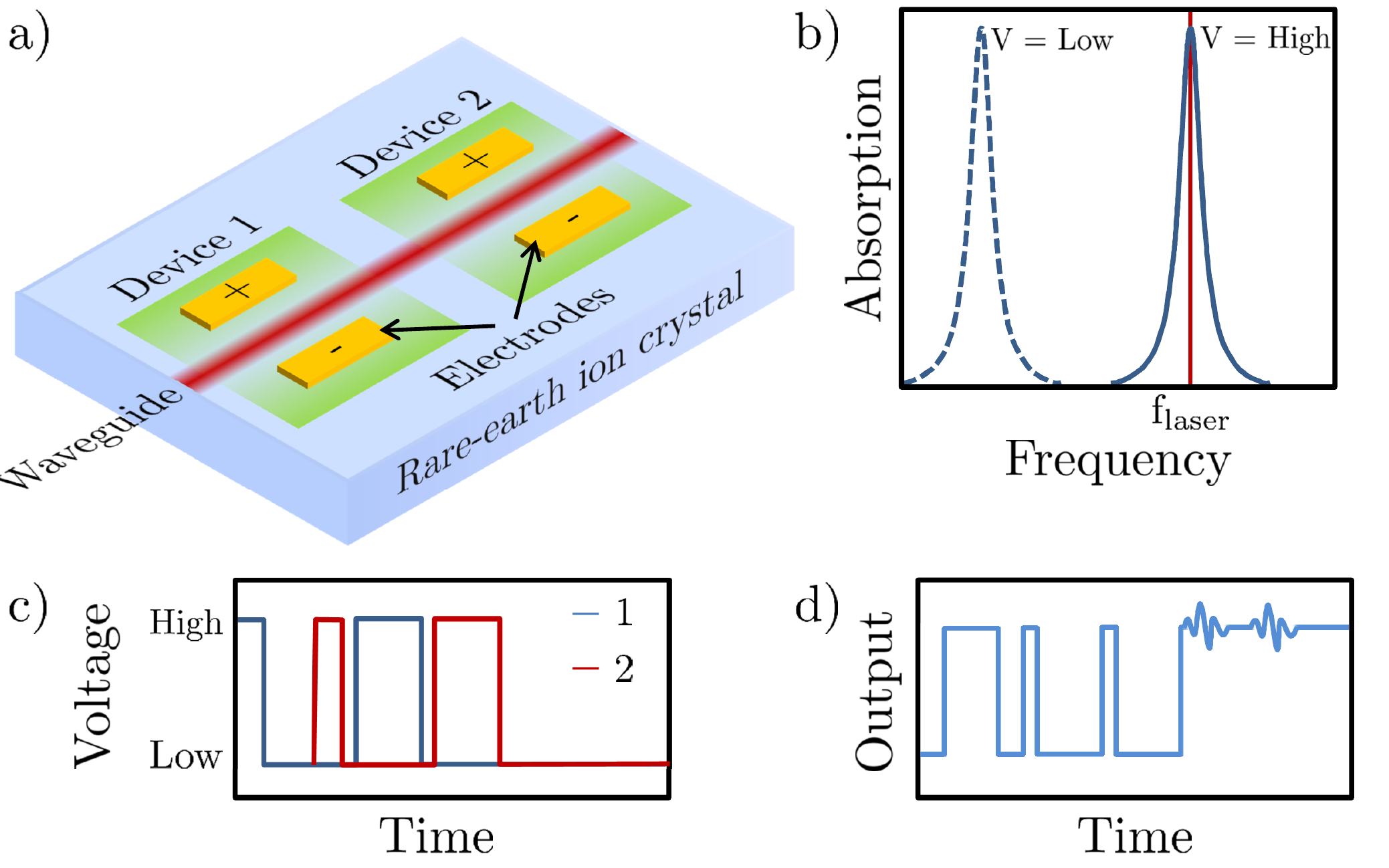}
\caption{A simple example of a device utilizing the proposed architecture for integrated rare-earth-ion quantum systems. Electrodes are incorporated with a waveguide-based platform to control individual components through the linear Stark shift of the ions. Two photon echo devices are shown in (a). When the voltage is switched from low to high the ions are shifted into resonance with the excitation laser frequency (b). Panels (c) and (d) show an example of the time evolution of the applied voltage and waveguide output for the two photon echo devices.}
\label{fig:arch}
\end{center}
\vspace{-0.5cm}
\end{figure}

The complex optical circuitry enabled by a waveguide-based platform offers great promise for constructing large, integrated rare-earth quantum systems. To capitalize on this potential we propose a planar waveguide architecture where the control of individual components
is achieved by Stark shifting optical transitions through the application of voltages to electrodes on the chip. The concept of the proposed architecture is illustrated in Fig.~\ref{fig:arch}. The fundamental idea is to create components by fast, local electronic manipulation of rare-earth-ion ensembles rather than changing the frequency of the light. The concept of utilizing a common source mode has been proposed for other architectures~\cite{Kane1998} and is suitable for rare-earth materials because the achievable Stark shifts are many orders of magnitude greater than the homogeneous linewidth of the optical transitions of interest. Furthermore, Stark shifting the ensemble frequency can be achieved rapidly and with low optical losses. 

The scheme allows multiple components to be concatenated such as the operation of two simple two-pulse photon echo devices in series as shown in Fig.~\ref{fig:arch}. The goal of this architecture is to combine non-classical sources, quantum memories, and control elements such as filters, modulators, and routers all on the one chip. The application of electric fields in rare-earth-ion materials can be controlled to perform coherent transients~\cite{Macfarlane2007}, spectral and temporal shaping of the ensemble~\cite{Beavan2013} and quantum protocols~\cite{Hedges2010}. For example, some of the earliest examples of photon echoes in rare-earth-ion crystals were achieved with a single frequency cw laser and using the Stark interaction to bring an ensemble in and out of resonance to create the required pulses~\cite{Genack1980}. With the configurability and fast switching available with modern micro-electrodes a particular waveguide region could be arbitrarily manipulated to form a quantum memory, a non-classical light source, a filter, a modulator, or to be completely non-interacting with the excitation laser frequency. Such an integrated chip would provide a physical implementation for engineering and maintaining a large resource of photonic qubits.

For the proposed architecture to be realizable it is necessary to develop planar waveguide structures incorporating rare-earth-ion dopants where the quantum coherence characteristics of the ions are preserved. Rare-earth-ion waveguide systems have been previously fabricated in LiNbO$_3$ using proton-exchange~\cite{Hastings-Simon2006} or Ti$^{4+}$ in-diffusion~\cite{Sinclair2010,Saglamyurek2011,Saglamyurek2014}. This approach combines the spectroscopic properties of rare-earth-ion dopants with the high performance waveguide technology that extends from the classical communications industry~\cite{Cusso2001}. A notable achievement of this research direction was the achievement of atomic frequency comb (AFC) storage of photon entanglement in a Ti$^{4+}$:Tm$^{3+}$:LiNbO$_3$ waveguide~\cite{Saglamyurek2011} and more recently an integrated processor for photonic quantum states~\cite{Saglamyurek2014}.

Although the use of existing waveguide technology provides thoroughly researched and developed fabrication methods, the spectroscopic properties of rare-earth ions in materials such as LiNbO$_3$ and KTP fall short of the properties available in other materials~\cite{Thiel2012}. For example, rare-earth dopants in LiNbO$_3$ demonstrate high levels of inhomogeneous broadening, spectral diffusion, and relatively short coherence times~\cite{Thiel2012}. Thus, it is not possible to engineer optimum components for quantum state creation or highly efficient, long term state storage. 

An alternate approach for constructing high performance rare-earth-ion waveguides is to grow single crystalline thin films on lower refractive index substrates to form slab waveguides. However, achieving the desired crystal quality in a sub-wavelength thickness film is yet to be demonstrated. For example, the 3~$\mu$m thick 2\% Eu$^{3+}$:Y$_2$O$_3$ film grown by Flinn et al. via metal-organic chemical vapor deposition possessed an inhomogeneous linewidth of $90 \pm 20$ GHz and a homogeneous linewidth of 12.1 MHz~\cite{Flinn1994}. Such results are a stark contrast from the 8 GHz inhomogeneous linewidth and 760 Hz homogeneous linewidth of a 2\% bulk sample of the same material~\cite{Macfarlane1981a}.

In this paper we perform initial demonstrations of a novel architecture for rare-earth-ion waveguides designed to preserve the properties available in high-quality single crystals. The technology is based on high refractive index glass waveguides that utilize bulk rare-earth-ion crystals as an active substrate. The optically active ions are then accessed by the evanescent field that extends into the substrate as light in the glass is guided by total internal reflection. By monitoring the output from the waveguide the inhomogeneous broadening, the homogeneous broadening and the spin state lifetime of ions within 100 nm of the crystal surface were measured. The ion properties reported in this paper and the ease with which waveguides can be fabricated using any one of the myriad of host crystals for rare-earth-ions establishes a solid platform for large scale integration of rare-earth quantum hardware.

\begin{figure}
\begin{center}
\includegraphics[width=0.45\textwidth]{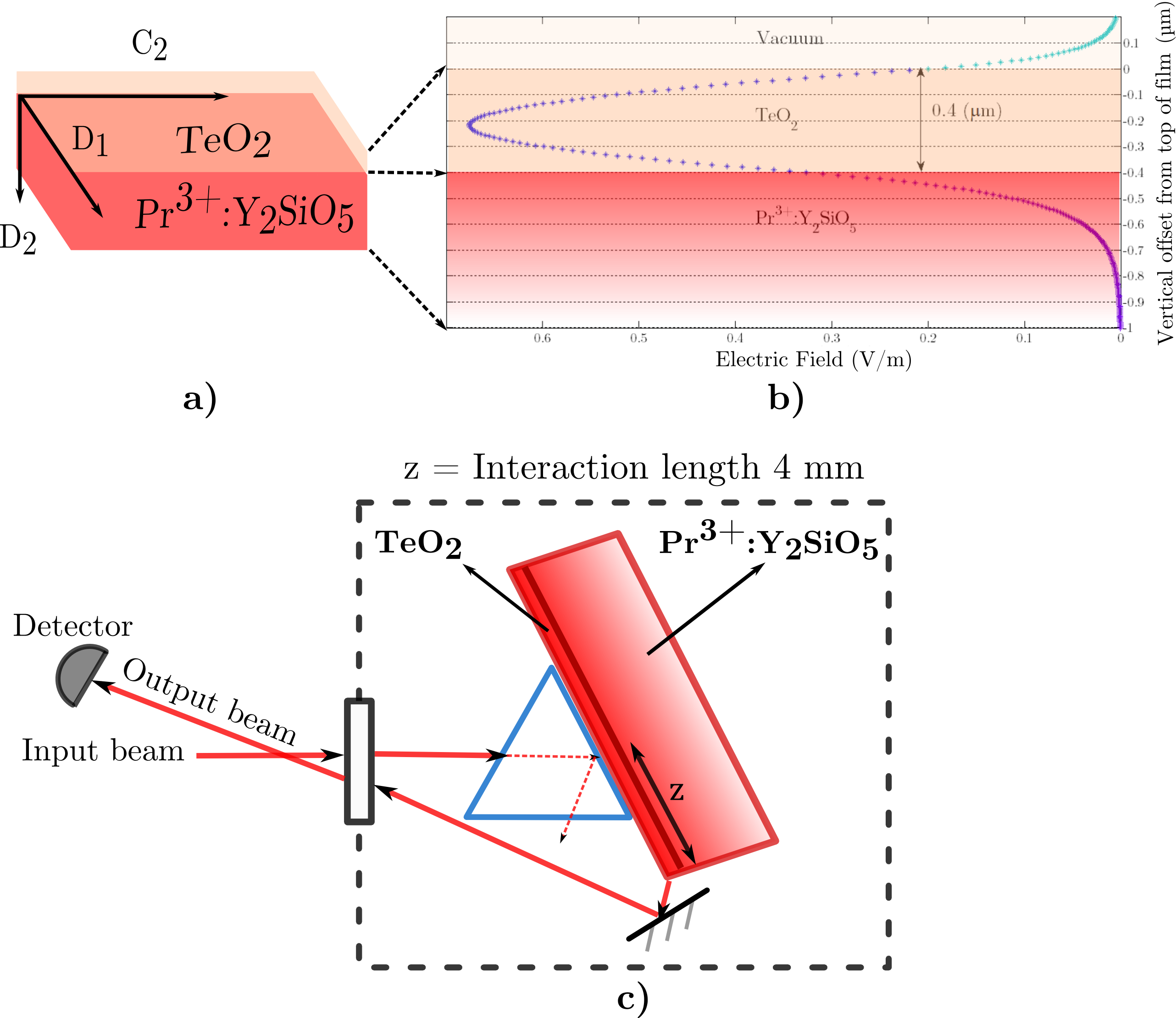}
\caption{(a) Crystalline axes relative to the film deposition. (b) The electric field amplitude E$_x$ of the fundamental TE mode supported by the 400~nm TeO$_{2}$ film waveguide on a Pr$^{3+}$:Y$_{2}$SiO$_{5}$ crystal substrate. (c) Experimental setup inside the cryostat used to perform the characterization of the TeO$_2$ on Pr$^{3+}$:Y$_2$SiO$_5$ waveguide.}
\label{fig:pr}
\end{center}
\end{figure}

The slab waveguide utilized in this work consisted of a $400 \pm 10$ nm thick TeO$_{2}$  thin film ($n_{\textrm{eff}} = 2.05$) on a 0.005\%\ Pr$^{3+}$:Y$_{2}$SiO$_{5}$ crystal ($n_{\textrm{sub}} = 1.806$). The glass film was formed by room temperature sputter deposition, which has been previously shown to produce high quality and low loss TeO$_2$ thin films~\cite{Madden2009}. For the polarization of the guided light to be parallel to the direction of maximum absorption ($\parallel$ D$_2$), the fundamental TE mode was excited in the film. Fig.~\ref{fig:pr} illustrates the calculated relative strength of the electric field propagating in the vacuum, the TeO$_2$ film, and in the Pr$^{3+}$:Y$_{2}$SiO$_{5}$ substrate. From calculations of the simulated waveguide mode, the evanescent field in the rare-earth-ion doped crystal substrate decays to its e$^{-1}$ point over 130~nm and the ratio of the power in the substrate to the total power in the guided mode is 7.2\%. 

%

\begin{figure}
\begin{center}
\includegraphics[width=0.5\textwidth, trim = 0.5cm 0.1cm 0.5cm 1cm, clip=true]{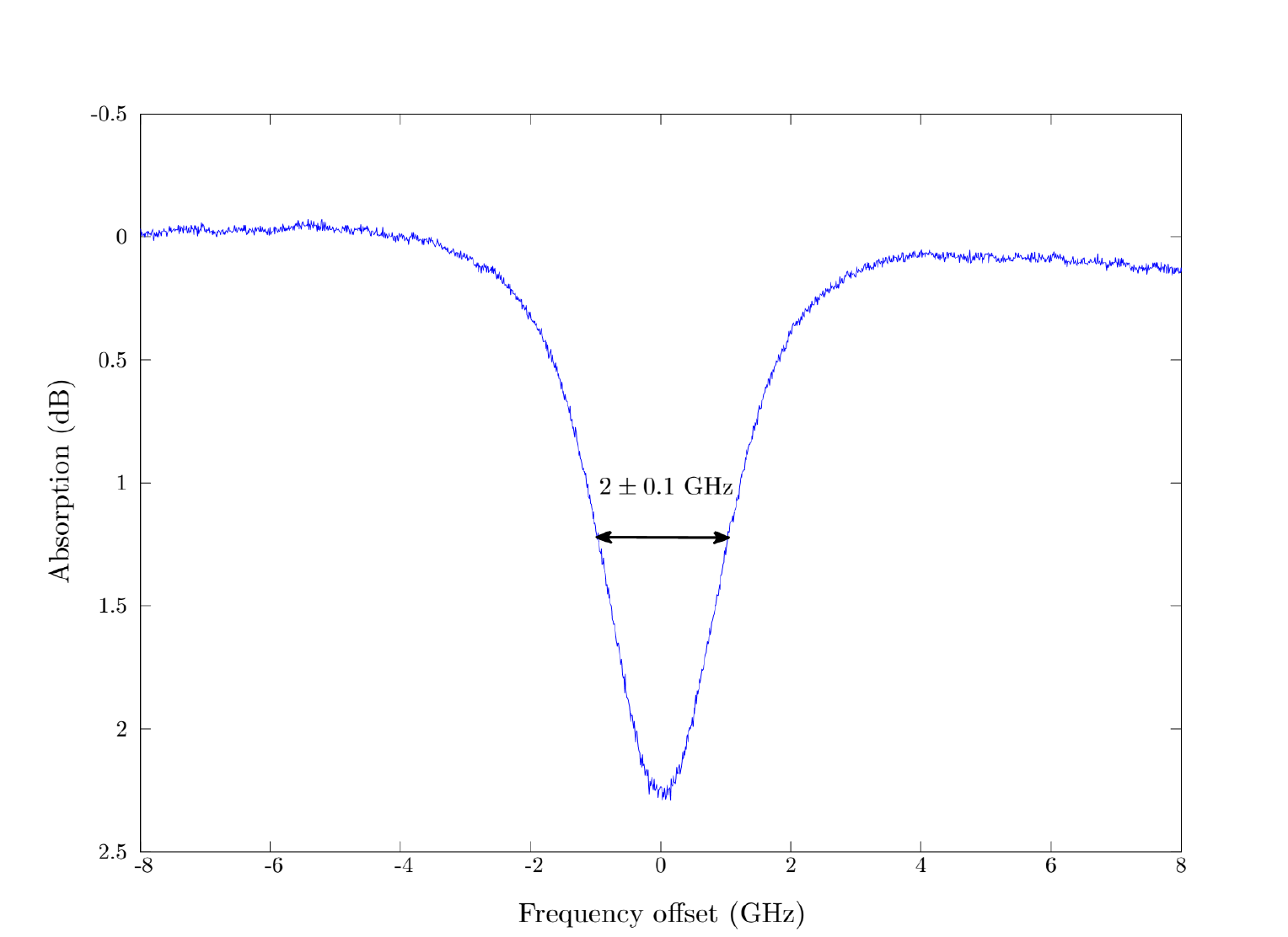}
\caption{The absorption spectrum of the 0.005\%~Pr$^{3+}$:Y$_{2}$SiO$_{5}$ substrate probed by the evanescent field from light guided in the thin TeO$_{2}$ film waveguide.}
\label{fig:Abs}
\end{center}
\end{figure}

\begin{figure}
\begin{center}
\includegraphics[width=0.5\textwidth, trim = 0.3cm 0.1cm 0.5cm 0.8cm, clip=true]{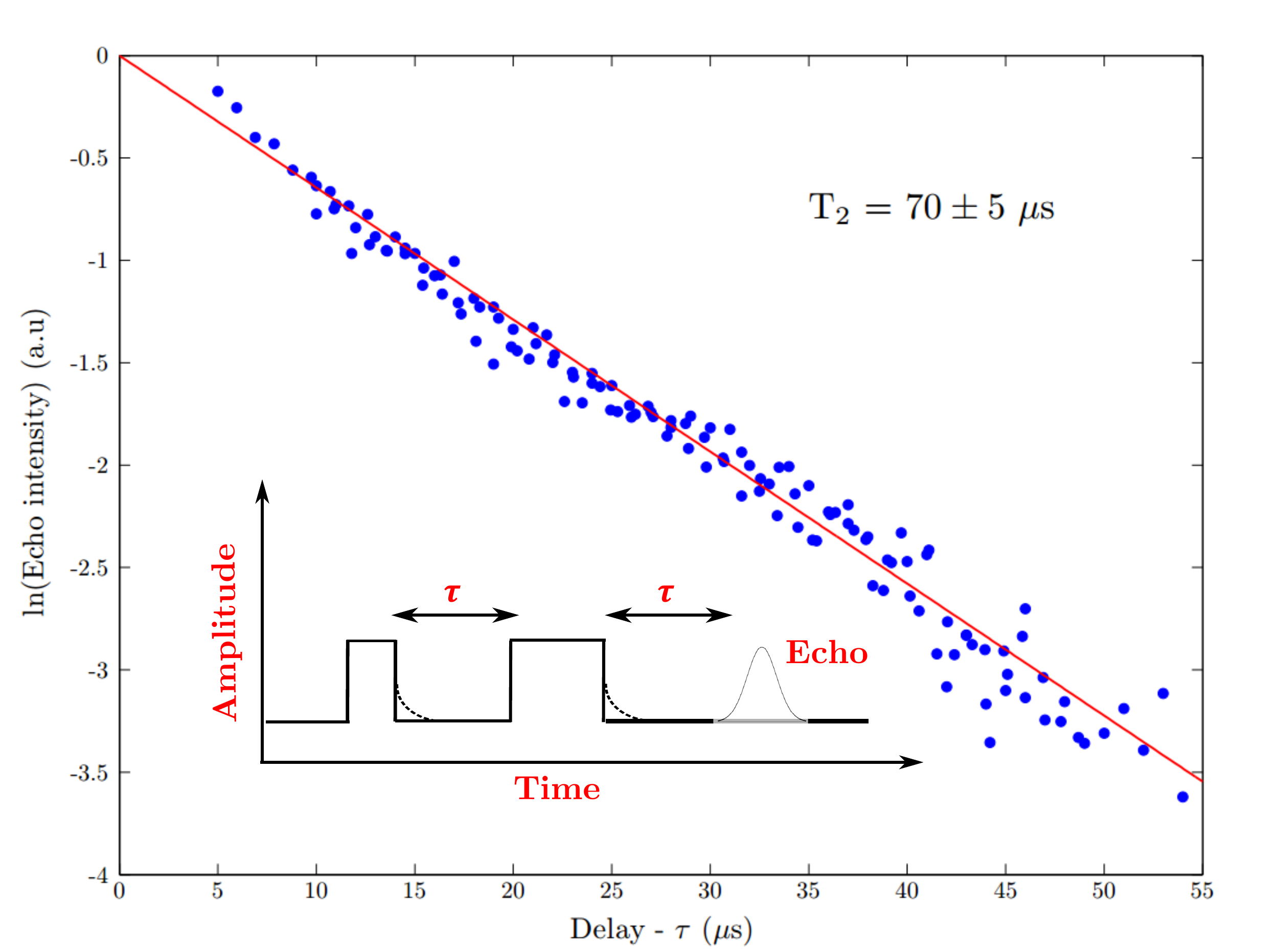}
\caption{Photon echo intensity as a function of the delay time $\tau$.}
\label{fig:delay}
\end{center}
\end{figure}

\begin{figure}
\begin{center}
\includegraphics[width=0.48\textwidth, trim = 0.5cm 0.1cm 0.5cm 0.5cm, clip=true]{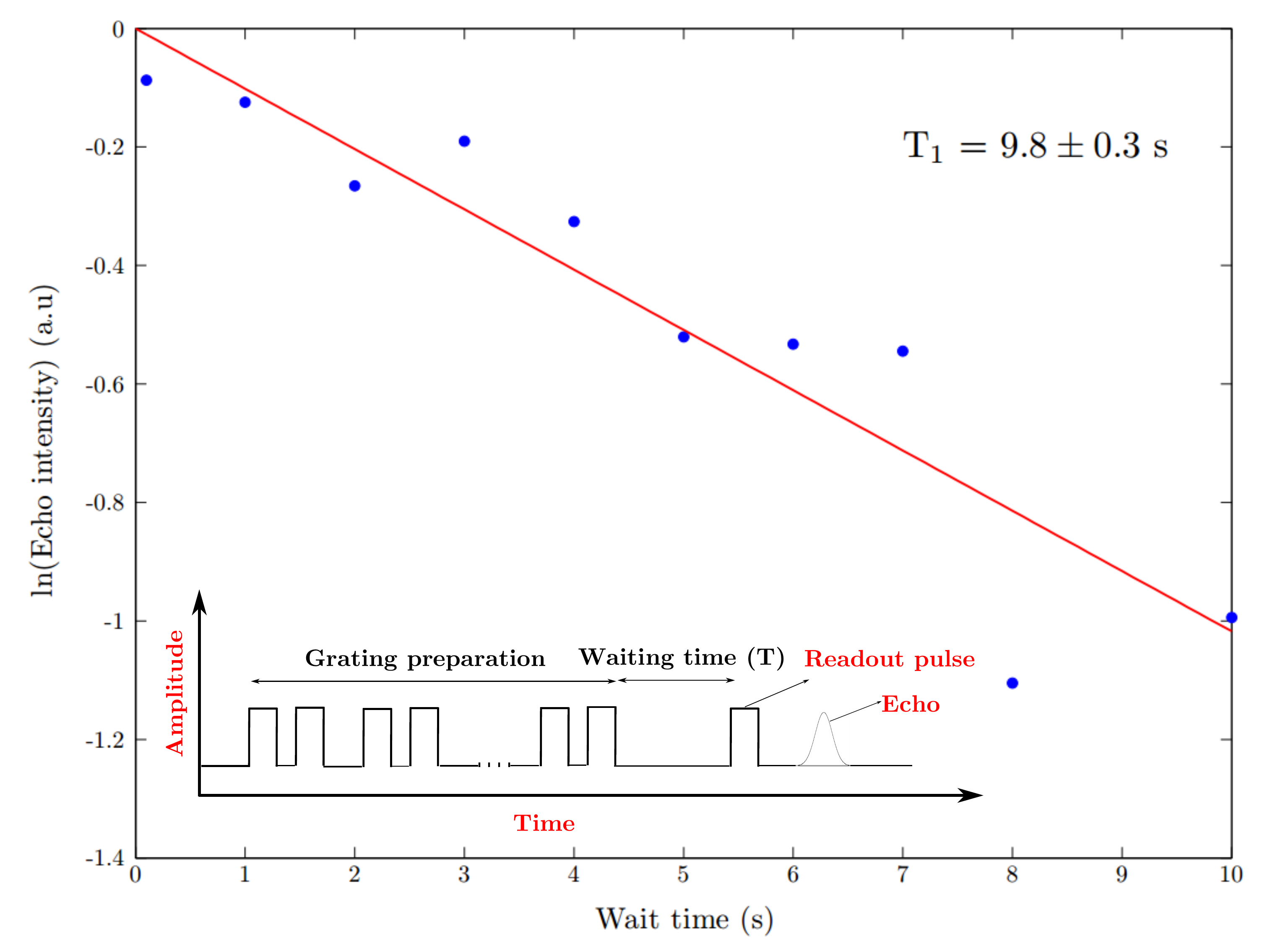}
\caption{The output echo intensity resultant from a pulse input onto a spectral grating as a function of the delay between writing and probing the grating. The measured decay time of the echo amplitude is representative of the lifetime (T$_1$) of the hyperfine spin states. }
\label{fig:spinecho}
\end{center}
\end{figure}

To probe the spectroscopic properties of the ions interacting with the waveguide mode, the experimental setup shown in Fig.~\ref{fig:pr} was used. A commercial Coherent $699-29$ dye laser with a linewidth on the order of 1~MHz was tuned to the $^3$H$_4 \leftrightarrow ^1$D$_2$ transition of Site 1 Pr$^{3+}$ ions at 605.977 nm. For pulsed measurements of coherence and spin state life times the excitation beam was gated with an acousto-optic modulator to create the photon echo sequences shown in Figs.~\ref{fig:delay} and ~\ref{fig:spinecho}. Light was then coupled in the planar slab waveguide using a rutile prism held in contact with the sample. In order to launch into the first TE mode in the waveguide, the angle of incidence of the beam on the air-prism interface was $18\pm0.1^{\circ}$. The optimized spot size at the prism-waveguide interface was observed to be approximately 80 $\mu$m, which allowed a coupling efficiency on the order of 45\%. The prism-waveguide assembly was placed in a helium bath cryostat and cooled to 2 K.
\\
\linebreak
The inhomogeneously broadened absorption line of an ensemble of ions is sensitive to the level of mechanical stress and static disorder in the volume of interest. Hence, the ensemble absorption provided an initial indication of the properties of the ions interacting with the waveguide mode relative to the well characterized properties of bulk ions. The inhomogeneous line was observed by scanning the laser 20 GHz and recording the resultant absorption spectrum, which is shown in Fig.~\ref{fig:Abs}. The inhomogeneously broadened line of evanescently coupled ions possessed a linewidth of $2\pm0.1$ GHz, which is consistent with the linewidth of bulk samples with the same Pr$^{3+}$ doping concentration~\cite{Equall1995}. The absence of excess broadening indicates that the film deposition process has not introduced addition static disorder into the crystal lattice. Furthermore, the narrow linewidth is strong evidence that the crystal field experienced by the ions and hence, their absorption properties, have not been significantly modified. The observed 2.25 dB of absorption over an interaction length of $4\pm1$ mm is consistent with the expected degree of absorption in the waveguide mode from the modeled mode profile assuming that ions at all depths are contributing. It should also be noted that the uncertainties are sufficiently large that it is possible that a region at the crystal surface could be present in which ions did not contribute to the observed signal. From our model, the upper bound on the depth of such a region was 16~nm.
 
The homogeneous linewidth of ions within an ensemble acts as a much more sensitive probe of the environment compared to the inhomogeneous linewidth. To probe the homogeneous linewidth of the ions in the waveguide mode, two-pulse photon echo experiments were performed. Due to laser power restrictions the excitation pulses used were longer than the laser coherence time (3~$\mu$s for the $\pi/2$-pulse and 6~$\mu$s for the $\pi$ pulse). However, a higher-resolution laser was not available at that time. Fig.~\ref{fig:delay} shows the averaged maximum intensity of the photon echo as a function of the delay $\tau$. The optical coherence time $T_{2}$ is $70 \pm 5\ \mu$s, which corresponds to a homogeneous linewidth of $4.5\pm0.3$~kHz. This is the narrowest linewidth achieved in a solid-state waveguide and is sufficient to allow demonstrations of non-classical light sources based on rephased amplified spontaneous emission and quantum memories based on either the gradient echo memory or AFC protocols.

We note that the homogeneous linewidth is approximately a factor of 1.6 broader than results obtained in zero-field bulk sample measurements in crystals of an equal Pr$^{3+}$ concentration under similar conditions~\cite{Equall1995}. The slight increase in broadening is attributed to a combination of three factors: laser instability, instantaneous spectral diffusion (ISD) and the impact of the guiding TeO$_2$ layer. Given the current measurements it is not possible to determine the precise contribution of each factor. Because the laser pulses exceed the laser coherence time there is a reduced degree of spectral overlap between the pulses that reduces the ability to effectively rephase the initial coherence of the ensemble. This effect also makes it difficult to predict the contribution of ISD because the excitation density of the ensemble is impacted. As a reference, assuming an excitation bandwidth of 1 MHz, the excitation density of Site 1 Pr$^{3+}$ ions in the substrate was $\rho_e = 4 \times 10^{14}$ cm$^{-3}$. Given the coefficient of ISD in Pr$^{3+}$:Y$_{2}$SiO$_{5}$ in Reference~\cite{Equall1995} ($1.2 \times 10^{-11}$ Hz/cm$^{3}$), 4.8 kHz of broadening is expected. Without knowledge of the broadening due to laser instability and ISD the impact of the guiding TeO$_2$ cannot be stated definitively. However, even if the factor of 1.6 increase in broadening was solely due to interactions between the ions and the glass film, the architecture would still allow the pursuit of integrated quantum systems using rare-earth ions.




To allow long term shelving of ions for protocols involving hole-burning or to achieve quantum state storage on the order of minutes to hours, the lifetime of the nuclear spin states observed in bulk crystals must be preserved in the waveguide architecture. To probe the ground state nuclear spin level lifetime, a stimulated echo measurement was performed. A series of 1000 $\pi/2$-pulse pairs created a spectral grating, which was then probed with a further $\pi/2$-pulse after a delay $T$. In this case a dye laser was available that was stabilized to a sub-kHz linewidth by locking to a temperature stabilized, ultra-low expansion optical cavity. The intensity of the three pulse echo is dependent upon the preservation of the spectral grating over time. The lifetime of the spin states provides an upper bound on the echo decay rate with $T$.

Fig.~\ref{fig:spinecho} shows the echo intensity as a function of the delay $T$, which ranged from 0.1 -- 1000 s. The decay behavior indicates the presence of short lived states, with a lifetime of $9.8\pm0.3$~s, and states with significantly longer lifetimes. Such behavior has been previously observed in holeburning studies of this material~\cite{Holliday1993}. The stimulated echo measurements indicate that the close proximity to the interface with the guiding glass medium has not significantly impacted the hyperfine relaxation rate compared to the bulk crystal. Furthermore, the observation of a well-defined echo at $T = 1000$ s is a strong indication that spectral diffusion is effectively absent even in this waveguide architecture. \\
 
To the best of the authors' knowledge, the characterization of the substrate ions performed here is the first use of an evanescent field to perform coherent transients in a solid state rare-earth-ion system. Furthermore, it is the first characterization of ions within one optical wavelength from a bulk crystal surface. The spectroscopic properties of the substrate ions probed by the guided mode are significantly different from the properties of previously characterized rare-earth-ion waveguide devices~\cite{Hastings-Simon2006,Sinclair2010}. By using the substrate as the active medium, the stress introduced into the host material by the waveguide fabrication process is minimized allowing narrow inhomogeneous linewidths to be achieved in a miniaturized platform. The minimization of inhomogeneity in rare-earth-ion crystals is essential to allow devices such as highly efficient, large bandwidth quantum memories with spin state storage~\cite{Hedges2010}. Furthermore, no evidence of spectral diffusion on a 15~minute time-scale was observed in the investigated slab waveguide. This is a stark contrast to waveguides fabricated in rare-earth-doped LiNbO$_3$, where rapid spectral diffusion limits spectral resolution to the 1~MHz level for 50~$\mu$s time scales~\cite{Sinclair2010}. Therefore, the waveguide structure demonstrated in this work constitutes an enabling architecture for large scale integrated rare-earth-ion quantum devices. 

Importantly, the architecture proposed and investigated here allows the optimization of the spectroscopic properties of the rare-earth crystal to be decoupled from the optimization of the waveguide fabrication. Thus, it is possible to capitalize on the cumulative knowledge of decades of high resolution spectroscopy by choosing a rare-earth-ion material that is optimally suited for quantum information processing applications. The deposition of a thin glass film of suitable quality and refractive index then allows a waveguide to be formed. 

An integrated rare-earth-ion crystal waveguide chip would be of significant interest not only to rare-earth-ion spectroscopy and applied quantum optics but to the broader quantum information processing community. Much attention is currently being devoted to waveguide based implementations of linear quantum optics to overcome the large overheads that prevent large scale implementations on optical tables. The waveguide architecture characterized in this paper accomplishes the goal of monolithic photonic circuits whilst combining the benefit of an on-chip quantum memory to synchronize numerous probabilistic events. Furthermore, triggerable single photon sources based on rephased amplified spontaneous emission~\cite{Ledingham2012} also promise high fidelity and efficiency, which would significantly reduce the heavy post-selection often incorporated into linear quantum optics experiments. Thus, the proposed waveguide architecture will be interesting as a system to achieve small scale quantum computation based on linear photonics.

This paper has proposed and demonstrated the feasibility of a novel waveguide architecture for rare-earth-ion materials. By guiding light in a thin, high refractive index glass deposited on the surface of a bulk rare-earth-ion crystal, the near surface ions can be probed by the evanescent field that extends into the substrate. The result is an effective waveguide, with little impact on the spectroscopic properties of the probed substrate ions. With the achievement of narrow inhomogeneous and homogeneous linewidths, negligible material related spectral diffusion and long hyperfine state lifetimes, future research will focus on fabricating and demonstrating waveguide-based devices such as gradient-echo quantum memories and single photon sources. The ultimate aim is then to combine such quantum components with other active and passive devices into a single, large scale integrated crystal-chip. The achievable photonic circuitry paired with the fast and precise control of the substrate ion ensembles with electric fields will allow new regimes of rare-earth-ion quantum optics to be probed in addition to facilitating research towards practical quantum communication and computing technology.
\newpage
\bibliography{library_mod}

\begin{thebibliography}{20}
\expandafter\ifx\csname natexlab\endcsname\relax\def\natexlab#1{#1}\fi
\expandafter\ifx\csname bibnamefont\endcsname\relax
  \def\bibnamefont#1{#1}\fi
\expandafter\ifx\csname bibfnamefont\endcsname\relax
  \def\bibfnamefont#1{#1}\fi
\expandafter\ifx\csname citenamefont\endcsname\relax
  \def\citenamefont#1{#1}\fi
\expandafter\ifx\csname url\endcsname\relax
  \def\url#1{\texttt{#1}}\fi
\expandafter\ifx\csname urlprefix\endcsname\relax\def\urlprefix{URL }\fi
\providecommand{\bibinfo}[2]{#2}
\providecommand{\eprint}[2][]{\url{#2}}

\bibitem[{\citenamefont{Hedges et~al.}(2010)\citenamefont{Hedges, Longdell, Li,
  and Sellars}}]{Hedges2010}
\bibinfo{author}{\bibfnamefont{M.~P.} \bibnamefont{Hedges}},
  \bibinfo{author}{\bibfnamefont{J.~J.} \bibnamefont{Longdell}},
  \bibinfo{author}{\bibfnamefont{Y.}~\bibnamefont{Li}}, \bibnamefont{and}
  \bibinfo{author}{\bibfnamefont{M.~J.} \bibnamefont{Sellars}},
  \bibinfo{journal}{Nature} \textbf{\bibinfo{volume}{465}}
  (\bibinfo{year}{2010}).

\bibitem[{\citenamefont{Jobez et~al.}(2014)\citenamefont{Jobez, Usmani,
  Timoney, Laplane, Gisin, and Afzelius}}]{Jobez2014}
\bibinfo{author}{\bibfnamefont{P.}~\bibnamefont{Jobez}},
  \bibinfo{author}{\bibfnamefont{I.}~\bibnamefont{Usmani}},
  \bibinfo{author}{\bibfnamefont{N.}~\bibnamefont{Timoney}},
  \bibinfo{author}{\bibfnamefont{C.}~\bibnamefont{Laplane}},
  \bibinfo{author}{\bibfnamefont{N.}~\bibnamefont{Gisin}}, \bibnamefont{and}
  \bibinfo{author}{\bibfnamefont{M.}~\bibnamefont{Afzelius}},
  \bibinfo{journal}{New Journal of Physics} \textbf{\bibinfo{volume}{16}}
  (\bibinfo{year}{2014}).

\bibitem[{\citenamefont{Ledingham et~al.}(2012)\citenamefont{Ledingham, Naylor,
  and Longdell}}]{Ledingham2012}
\bibinfo{author}{\bibfnamefont{P.~M.} \bibnamefont{Ledingham}},
  \bibinfo{author}{\bibfnamefont{W.~R.} \bibnamefont{Naylor}},
  \bibnamefont{and} \bibinfo{author}{\bibfnamefont{J.~J.}
  \bibnamefont{Longdell}}, \bibinfo{journal}{Physical Review Letters}
  \textbf{\bibinfo{volume}{109}} (\bibinfo{year}{2012}).

\bibitem[{\citenamefont{Beavan et~al.}(2013)\citenamefont{Beavan, Goldschmidt,
  and Sellars}}]{Beavan2013}
\bibinfo{author}{\bibfnamefont{S.~E.} \bibnamefont{Beavan}},
  \bibinfo{author}{\bibfnamefont{E.~A.} \bibnamefont{Goldschmidt}},
  \bibnamefont{and} \bibinfo{author}{\bibfnamefont{M.~J.}
  \bibnamefont{Sellars}}, \bibinfo{journal}{Journal of the Optical Society of
  America B} \textbf{\bibinfo{volume}{30}} (\bibinfo{year}{2013}).

\bibitem[{\citenamefont{Tay et~al.}(2010)\citenamefont{Tay, Ledingham, and
  Longdell}}]{Tay2010}
\bibinfo{author}{\bibfnamefont{J.~W.} \bibnamefont{Tay}},
  \bibinfo{author}{\bibfnamefont{P.~M.} \bibnamefont{Ledingham}},
  \bibnamefont{and} \bibinfo{author}{\bibfnamefont{J.~J.}
  \bibnamefont{Longdell}}, \bibinfo{journal}{Appl. Opt.}
  \textbf{\bibinfo{volume}{49}} (\bibinfo{year}{2010}).

\bibitem[{\citenamefont{Saglamyurek et~al.}(2014)\citenamefont{Saglamyurek,
  Sinclair, Slater, Heshami, Oblak, and Tittel}}]{Saglamyurek2014}
\bibinfo{author}{\bibfnamefont{E.}~\bibnamefont{Saglamyurek}},
  \bibinfo{author}{\bibfnamefont{N.}~\bibnamefont{Sinclair}},
  \bibinfo{author}{\bibfnamefont{J.~A.} \bibnamefont{Slater}},
  \bibinfo{author}{\bibfnamefont{K.}~\bibnamefont{Heshami}},
  \bibinfo{author}{\bibfnamefont{D.}~\bibnamefont{Oblak}}, \bibnamefont{and}
  \bibinfo{author}{\bibfnamefont{W.}~\bibnamefont{Tittel}},
  \bibinfo{journal}{New Journal of Physics} \textbf{\bibinfo{volume}{16}}
  (\bibinfo{year}{2014}).

\bibitem[{\citenamefont{Wang et~al.}(2009)\citenamefont{Wang, Fan, Wang, Du,
  Zhang, Kang, Jiang, Wu, and Gao}}]{Wang2009b}
\bibinfo{author}{\bibfnamefont{H.-H.} \bibnamefont{Wang}},
  \bibinfo{author}{\bibfnamefont{Y.-F.} \bibnamefont{Fan}},
  \bibinfo{author}{\bibfnamefont{R.}~\bibnamefont{Wang}},
  \bibinfo{author}{\bibfnamefont{D.-M.} \bibnamefont{Du}},
  \bibinfo{author}{\bibfnamefont{X.-J.} \bibnamefont{Zhang}},
  \bibinfo{author}{\bibfnamefont{Z.-H.} \bibnamefont{Kang}},
  \bibinfo{author}{\bibfnamefont{Y.}~\bibnamefont{Jiang}},
  \bibinfo{author}{\bibfnamefont{J.-H.} \bibnamefont{Wu}}, \bibnamefont{and}
  \bibinfo{author}{\bibfnamefont{J.-Y.} \bibnamefont{Gao}},
  \bibinfo{journal}{Optics Express} \textbf{\bibinfo{volume}{17}}
  (\bibinfo{year}{2009}).

\bibitem[{\citenamefont{Kane}(1998)}]{Kane1998}
\bibinfo{author}{\bibfnamefont{B.~E.} \bibnamefont{Kane}},
  \bibinfo{journal}{Nature} \textbf{\bibinfo{volume}{393}}
  (\bibinfo{year}{1998}).

\bibitem[{\citenamefont{Macfarlane}(2007)}]{Macfarlane2007}
\bibinfo{author}{\bibfnamefont{R.~M.} \bibnamefont{Macfarlane}},
  \bibinfo{journal}{Journal of Luminescence} \textbf{\bibinfo{volume}{125}}
  (\bibinfo{year}{2007}).

\bibitem[{\citenamefont{Genack et~al.}(1980)\citenamefont{Genack, Weitz,
  Macfarlane, Shelby, and Schenzle}}]{Genack1980}
\bibinfo{author}{\bibfnamefont{A.}~\bibnamefont{Genack}},
  \bibinfo{author}{\bibfnamefont{D.}~\bibnamefont{Weitz}},
  \bibinfo{author}{\bibfnamefont{R.}~\bibnamefont{Macfarlane}},
  \bibinfo{author}{\bibfnamefont{R.}~\bibnamefont{Shelby}}, \bibnamefont{and}
  \bibinfo{author}{\bibfnamefont{A.}~\bibnamefont{Schenzle}},
  \bibinfo{journal}{Physical Review Letters} \textbf{\bibinfo{volume}{45}}
  (\bibinfo{year}{1980}).

\bibitem[{\citenamefont{Hastings-Simon
  et~al.}(2006)\citenamefont{Hastings-Simon, Staudt, Afzelius, Baldi, Jaccard,
  Tittel, and Gisin}}]{Hastings-Simon2006}
\bibinfo{author}{\bibfnamefont{S.~R.} \bibnamefont{Hastings-Simon}},
  \bibinfo{author}{\bibfnamefont{M.~U.} \bibnamefont{Staudt}},
  \bibinfo{author}{\bibfnamefont{M.}~\bibnamefont{Afzelius}},
  \bibinfo{author}{\bibfnamefont{P.}~\bibnamefont{Baldi}},
  \bibinfo{author}{\bibfnamefont{D.}~\bibnamefont{Jaccard}},
  \bibinfo{author}{\bibfnamefont{W.}~\bibnamefont{Tittel}}, \bibnamefont{and}
  \bibinfo{author}{\bibfnamefont{N.}~\bibnamefont{Gisin}},
  \bibinfo{journal}{Optics Communications} \textbf{\bibinfo{volume}{266}}
  (\bibinfo{year}{2006}).

\bibitem[{\citenamefont{Sinclair et~al.}(2010)\citenamefont{Sinclair,
  Saglamyurek, George, Ricken, {La Mela}, Sohler, and Tittel}}]{Sinclair2010}
\bibinfo{author}{\bibfnamefont{N.}~\bibnamefont{Sinclair}},
  \bibinfo{author}{\bibfnamefont{E.}~\bibnamefont{Saglamyurek}},
  \bibinfo{author}{\bibfnamefont{M.}~\bibnamefont{George}},
  \bibinfo{author}{\bibfnamefont{R.}~\bibnamefont{Ricken}},
  \bibinfo{author}{\bibfnamefont{C.}~\bibnamefont{{La Mela}}},
  \bibinfo{author}{\bibfnamefont{W.}~\bibnamefont{Sohler}}, \bibnamefont{and}
  \bibinfo{author}{\bibfnamefont{W.}~\bibnamefont{Tittel}},
  \bibinfo{journal}{Journal of Luminescence} \textbf{\bibinfo{volume}{130}}
  (\bibinfo{year}{2010}).

\bibitem[{\citenamefont{Saglamyurek et~al.}(2011)\citenamefont{Saglamyurek,
  Sinclair, Jin, Slater, Oblak, Bussi\`{e}res, George, Ricken, Sohler, and
  Tittel}}]{Saglamyurek2011}
\bibinfo{author}{\bibfnamefont{E.}~\bibnamefont{Saglamyurek}},
  \bibinfo{author}{\bibfnamefont{N.}~\bibnamefont{Sinclair}},
  \bibinfo{author}{\bibfnamefont{J.}~\bibnamefont{Jin}},
  \bibinfo{author}{\bibfnamefont{J.~A.} \bibnamefont{Slater}},
  \bibinfo{author}{\bibfnamefont{D.}~\bibnamefont{Oblak}},
  \bibinfo{author}{\bibfnamefont{F.}~\bibnamefont{Bussi\`{e}res}},
  \bibinfo{author}{\bibfnamefont{M.}~\bibnamefont{George}},
  \bibinfo{author}{\bibfnamefont{R.}~\bibnamefont{Ricken}},
  \bibinfo{author}{\bibfnamefont{W.}~\bibnamefont{Sohler}}, \bibnamefont{and}
  \bibinfo{author}{\bibfnamefont{W.}~\bibnamefont{Tittel}},
  \bibinfo{journal}{Nature} \textbf{\bibinfo{volume}{469}}
  (\bibinfo{year}{2011}).

\bibitem[{\citenamefont{Cuss\'{o} et~al.}(2001)\citenamefont{Cuss\'{o},
  Lifante, Mu$\tilde{\textrm{n}}$toz, Cantelar, Nevado, Cino, {De Micheli}, and
  Sohler}}]{Cusso2001}
\bibinfo{author}{\bibfnamefont{F.}~\bibnamefont{Cuss\'{o}}},
  \bibinfo{author}{\bibfnamefont{G.}~\bibnamefont{Lifante}},
  \bibinfo{author}{\bibfnamefont{J.~A.}
  \bibnamefont{Mu$\tilde{\textrm{n}}$toz}},
  \bibinfo{author}{\bibfnamefont{E.}~\bibnamefont{Cantelar}},
  \bibinfo{author}{\bibfnamefont{R.}~\bibnamefont{Nevado}},
  \bibinfo{author}{\bibfnamefont{A.}~\bibnamefont{Cino}},
  \bibinfo{author}{\bibfnamefont{M.~P.} \bibnamefont{{De Micheli}}},
  \bibnamefont{and} \bibinfo{author}{\bibfnamefont{W.}~\bibnamefont{Sohler}},
  \bibinfo{journal}{Radiation Effects and Defects in Solids}
  \textbf{\bibinfo{volume}{155}} (\bibinfo{year}{2001}).

\bibitem[{\citenamefont{Thiel et~al.}(2012)\citenamefont{Thiel, Sun,
  Macfarlane, B\"{o}ttger, and Cone}}]{Thiel2012}
\bibinfo{author}{\bibfnamefont{C.~W.} \bibnamefont{Thiel}},
  \bibinfo{author}{\bibfnamefont{Y.}~\bibnamefont{Sun}},
  \bibinfo{author}{\bibfnamefont{R.~M.} \bibnamefont{Macfarlane}},
  \bibinfo{author}{\bibfnamefont{T.}~\bibnamefont{B\"{o}ttger}},
  \bibnamefont{and} \bibinfo{author}{\bibfnamefont{R.~L.} \bibnamefont{Cone}},
  \bibinfo{journal}{Journal of Physics B: Atomic, Molecular and Optical
  Physics} \textbf{\bibinfo{volume}{45}} (\bibinfo{year}{2012}).

\bibitem[{\citenamefont{Flinn et~al.}(1994)\citenamefont{Flinn, Jang, Ganem,
  Jones, Meltzer, and Macfarlane}}]{Flinn1994}
\bibinfo{author}{\bibfnamefont{G.~P.} \bibnamefont{Flinn}},
  \bibinfo{author}{\bibfnamefont{K.~W.} \bibnamefont{Jang}},
  \bibinfo{author}{\bibfnamefont{J.}~\bibnamefont{Ganem}},
  \bibinfo{author}{\bibfnamefont{M.~L.} \bibnamefont{Jones}},
  \bibinfo{author}{\bibfnamefont{R.~S.} \bibnamefont{Meltzer}},
  \bibnamefont{and} \bibinfo{author}{\bibfnamefont{R.~M.}
  \bibnamefont{Macfarlane}}, \bibinfo{journal}{Journal of Luminescence}
  \textbf{\bibinfo{volume}{58}} (\bibinfo{year}{1994}).

\bibitem[{\citenamefont{Macfarlane and Shelby}(1981)}]{Macfarlane1981a}
\bibinfo{author}{\bibfnamefont{R.~M.} \bibnamefont{Macfarlane}}
  \bibnamefont{and} \bibinfo{author}{\bibfnamefont{R.~M.}
  \bibnamefont{Shelby}}, \bibinfo{journal}{Optics Communications}
  \textbf{\bibinfo{volume}{39}} (\bibinfo{year}{1981}).

\bibitem[{\citenamefont{Madden and Vu}(2009)}]{Madden2009}
\bibinfo{author}{\bibfnamefont{S.~J.} \bibnamefont{Madden}} \bibnamefont{and}
  \bibinfo{author}{\bibfnamefont{K.~T.} \bibnamefont{Vu}},
  \bibinfo{journal}{Opt. Express} \textbf{\bibinfo{volume}{17}}
  (\bibinfo{year}{2009}).

\bibitem[{\citenamefont{Equall et~al.}(1995)\citenamefont{Equall, Cone, and
  Macfarlane}}]{Equall1995}
\bibinfo{author}{\bibfnamefont{R.~W.} \bibnamefont{Equall}},
  \bibinfo{author}{\bibfnamefont{R.~L.} \bibnamefont{Cone}}, \bibnamefont{and}
  \bibinfo{author}{\bibfnamefont{R.~M.} \bibnamefont{Macfarlane}},
  \bibinfo{journal}{Physical Review B} \textbf{\bibinfo{volume}{52}}
  (\bibinfo{year}{1995}).

\bibitem[{\citenamefont{Holliday et~al.}(1993)\citenamefont{Holliday, Croci,
  Vauthey, and Wild}}]{Holliday1993}
\bibinfo{author}{\bibfnamefont{K.}~\bibnamefont{Holliday}},
  \bibinfo{author}{\bibfnamefont{M.}~\bibnamefont{Croci}},
  \bibinfo{author}{\bibfnamefont{E.}~\bibnamefont{Vauthey}}, \bibnamefont{and}
  \bibinfo{author}{\bibfnamefont{U.~P.} \bibnamefont{Wild}},
  \bibinfo{journal}{Physical Review B} \textbf{\bibinfo{volume}{47}}
  (\bibinfo{year}{1993}).

\end{thebibliography}


\begin{thebibliography}{99}
%
\bibitem {Jevon}
J. Longdell, M.Sellars and N. Manson, ``Demonstration of conditional quantum phase shift between ions in a solid,'' Phys. Rev. Lett. 93, (2004).

\bibitem {Morgan} 
M. Hedges, J. Longdell, Y. Li, and M. Sellars, ``Efficient quantum memory for light'', Nature 465, (2010). 

\bibitem{Clausen}
C. Clausen, I. Usmani, F. Bussieres, N. Sangouard, M. Afzelius, H. de Riedmatten and N. Gisin, ``Quantum storage of photonic entanglement in a crystal'', Nature 469, (2011)

\bibitem{Usmani}
I. Usmani, C. Clausen, F. Bussieres, N. Sangouard, M. Afzelius and N. Gisin, ``Heralded quantum entanglement between two crystals'', Nature Photonics 6, (2012)

\bibitem{Zhong2013}
Grace's paper

\bibitem{Patrick2010}
P.M. Ledingham, W.R. Naylor, J.J. Longdell, S.E. Beavan, M.J. Sellars, "Nonclassical photon streams using rephased amplified spontaneous emission", Phys. Rev. A, (2010)

\bibitem{Patrick}
P.M. Ledingham, W. Naylor, J. Longdell,``Experimental realization of light with time separated correlations by rephasing amplified spontaneous emission'', Phys. Rev. Lett. 109, (2012) 

\bibitem{Sarah}
S.E. Beavan, E.A. Goldschmidt, and M.J. Sellars, "Demonstration of a dynamic bandpass frequency filter in a rare-earth ion-doped crystal," J. Opt. Soc. Am. B 30, 1173-1177 (2013) 

\bibitem{Beavan}
S.E. Beavan, M. Hedges, M.J. Sellars,"Demonstration of Photon-Echo Rephasing of Spontaneous Emission", Phys. Rev. Lett. 109, (2012)

\bibitem{Ohlsson2002}
N. Ohlsson, R.K. Mohan, S. Kr\"{o}ll,  Quantum computer hardware based on rare-earth-ion-doped inorganic crystals, Opt. Commun. 201, 71--77 (2002)

\bibitem{Longdell}
J.J. Longdell, M.J. Sellars, ``Experimental demonstration of quantum-state tomography and qubit-qubit interactions for rare-earth-metal-ion-based solid-state qubits'', Phys. Rev. A, (2004)

\bibitem{Longdella}
J.J. Longdell, M.J. Sellars, N.B. Manson , ``Demonstration of Conditional Quantum Phase Shift Between Ions in a Solid'', Phys. Rev. Lett., (2004)

\bibitem{AhlefeldtThesis}
R. Ahlefeldt, ``Evaluation of a stoichiometric rare earth crystal for quantum computing'', PhD Thesis, Australian National University, (2013)

\bibitem{Sohler}
W. Sohler, H. Herrmann, R. Ricken, V. Quiring, M. George, S. Pal, X. Yang, K. H. Luo, C. Silberhorn, F. Kaiser, S. Tanzilli, E. Saglamyurek, N. Sinclair, D. Oblak, W. Tittel, "Integrated Photon Pair Sources, Quantum Memories, and Lasers in Lithium Niobate," in International Photonics and Optoelectronics Meetings, OSA Technical Digest (online) (Optical Society of America, 2012), paper IF1A.1 

\bibitem{Wolfgang}
E. Saglamyurek, N. Sinclair, J. Jin, J. A. Slater, D. Oblak, F. Bussieres, M. George, R. Ricken, W. Sohler and W. Tittel, ``Broadband waveguide quantum memory for
entangled photons'', Nature 469, (2011)

\bibitem{khu}
S.J. Madden, K.T. Vu, "Very low loss reactively ion etched Tellurium Dioxide planar rib waveguides for linear and non-linear optics", Optics Express, Vol. 17, Issue 20, pp. 17645-17651 (2009)

\bibitem{Equall}
R. Equall, R. Cone, R. Macfarlane, ``Homogeneous broadening and hyperfine structure of optical transitions in  Pr$^{3+}$:Y$_{2}$SiO$_{5}$'', Phys. Rev. B 52, (1995)

\bibitem{MorganThesis}
M.Hedges, "High performance solid state quantum memory", PhD Thesis, Australian National University, (2011)

\bibitem{Holliday}
K. Holliday, M. Croci, E. Vauthey, U.P. Wild, ``Spectral hole burning and holography in an Y$_{2}$SiO$_{5}$:Pr$^{3+}$ crystal'', Phys. Rev. B,Vol 47, Issue 22,(1993)

\end{thebibliography}
\newpage

\end{document}